\documentclass[prb, amsmath,amssymb, reprint]{revtex4-1}
\usepackage{graphicx}
\usepackage{mathptmx}

\begin{document}

\title{
Closed-loop electric currents and non-local resistance measurements \\ with wide F/I/N tunnel contacts
}

\author{Ya.\ B. Bazaliy}
  \affiliation{University of South Carolina, Columbia SC 29208, USA}
  \email{yar@physics.sc.edu}
\author{R. R. Ramazashvili}
 \affiliation{Laboratoire de Physique Th\'eorique, Universit\'e de Toulouse, CNRS, UPS, France}
 \email{revaz@irsamc.ups-tlse.fr}

\date{\today}

\begin{abstract}
Lateral spin valves are used to generate and characterize pure spin currents. Non-local voltage measured in such structures provides information about spin polarization and spin decay rates. For wide high-transparency F/N contacts it was shown that the Johnson-Silsbee non-local effect is substantially enriched by closed-loop electric currents driven by local spin injection in the electrically dangling part of the valve. For valves with low-transparency F/I/N tunnel contacts such circular currents are strongly suppressed, yet we show that the voltage modifications persist, may be significant, and must be accounted for in the data analysis.
\end{abstract}

\maketitle

\section{Introduction}
\label{sec:Introduction}

\begin{figure}[b]
\center
\includegraphics[width = 0.4\textwidth]{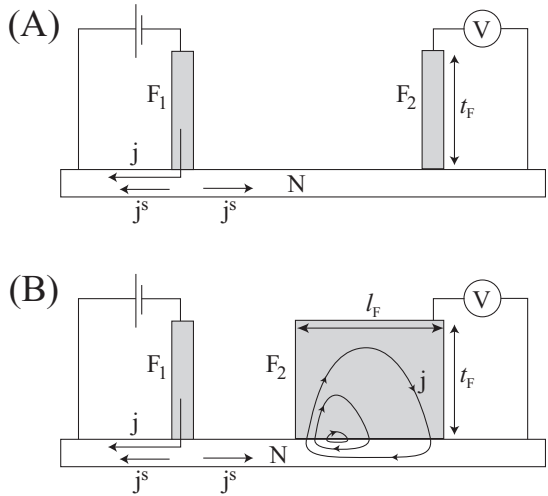}
    \caption{Non-local spin valve (NLSV) diagram. (A) Narrow measuring  contact, no electric current. (B) Wide measuring contact with electric current vortices.}
 \label{fig:nlsv}
\end{figure}

A non-local spin valve (NLSV) consists of a normal metal (N) line with two ferromagnetic (F) contacts (Fig. \ref{fig:nlsv}). The left contact F$_1$ injects spin-polarized electrons into N, where they diffuse away from the injection point, producing spin currents $j^s$ both in the left and right directions. At the same time, electric current $j$ cannot enter the electrically dangling part of the circuit on the right---consequently, only spin current is present there.\cite{johnson-silsbee_prl1985, johnson-silsbee_prb1987, johnson-silsbee_prb2007} The non-equilibrium electron state driven by such an injection of spin current results in a non-zero vol\-ta\-ge $V$, measured by an ideal voltmeter between the F$_2$ and $N$. The absence of $j$ in the electrically dangling part of the valve dictates a very general relation between $V$ and the spin accumulation beneath the contact F$_2$, known as the Johnson-Silsbee formula.\cite{johnson-silsbee_prl1985} The voltage $V$ turns out to be independent of the voltmeter probe positions as long as the thickness $t_F$ of the measuring electrode F$_2$ remains much larger than the spin diffusion length in that material [see Eqs. (\ref{eq:JS_ohmic}) and (\ref{eq:JS_tunnel})].

However, it was noticed \cite{bazaliy:apl2017} that the $j = 0$ condition holds only for narrow F$_2$ contacts (Fig. \ref{fig:nlsv}A). As the width $l_{F}$ exceeds the scale of appreciable variation of spin accumulation (Fig. \ref{fig:nlsv}B), closed-loop electric currents $j \neq 0$ develop, for\-ming a vortex centered at the F$_2$/N interface. No electric current enters or leaves the electrically dangling (right-hand) part of the device, all current loops are fully contained within it. Crucially, these current loops significantly suppress the measured vol\-ta\-ge $V$ and, generally, lead to its dependence on the voltmeter probe positions.

Conclusions of Ref.~\onlinecite{bazaliy:apl2017} were reached under the assumption of fully transparent interface between N and F$_2$, well satisfied in many realizations of spin valves. At the same time, certain N materials require a tunnel contact for spin injection and detection to overcome the con\-duc\-ti\-vi\-ty mismatch problem.\cite{schmidt_prbrc2000, rashba_prbrc2000} As a result, tunnel contact measurements more and more become the method of choice.\cite{joner_naturephys2007, dash_nature2009, tran_prl2009, li_naturecomm2011, han_naturecomm2013, vanterve_naturecomm2015, leutenatsmeyer_prl2018, drogeler_nanolett2016, gurram_natcomm2017, dankert_naturecomm2017, avsar_naturephys2017, spiesser_apl2019} Will they be affected by the closed-loop electric currents? As shown in Ref.~\onlinecite{bazaliy:apl2017}, each loop of induced current crosses the F$_2$/N interface (Fig. \ref{fig:nlsv}B) and, in the presence of a tunnel contact, such a current would have to flow across the highly resistive barrier. Na\"ively, one would expect a dramatic suppression of such currents by the tunnel barrier, and hence a recovery of the Johnson-Silsbee result.\cite{johnson-silsbee_prl1985} Below we show that such a conclusion is, in fact, incorrect: while the current does decrease with increasing tunnel resistance, nevertheless it significantly suppresses the measured non-local voltage, which may become substantially smaller than the Johnson-Silsbee value.

\section{Non-local voltage calculation}
\label{sec:main}

\subsection{Description of electric and spin transport}
We consider electric and spin currents in the diffusive regime, and assume collinear magnetizations of the injector F$_1$ and detector F$_2$ electrodes, as is the case in many NLSV measurements.  Transport is described by the Valet-Fert equations \cite{campbell:1967,valet-fert_prb1993,rashba_epjb2002, takahashi_prb2003}
in the notations of Ref.~\onlinecite{bazaliy:apl2017} (see Supplement).

Particle- and spin-current densities $j^{e}$,$j^{s}$ in the bulk are induced by the gradients of electrochemical and spin potentials $\mu$, $\mu^{s}$ and obey material equations
\begin{eqnarray}
\label{eq:j}
j^{e}_i &=& - \frac{\sigma}{e^2} \, (\nabla_i \mu + \frac{p}{2}\nabla_i \mu^s)
\\
\label{eq:js}
j^s_i &=& -\frac{\sigma}{2 e^2} \, (\nabla_i \mu^s + 2 p \nabla_i \mu)
\end{eqnarray}
where $\sigma = \sigma_{\uparrow} + \sigma_{\downarrow}$ is the conductivity of the material and $p = (\sigma_{\uparrow}-\sigma_{\downarrow})/\sigma$ is the ``current spin polarization'', present in F only. The spin quantization axis is chosen along the magnetization.

Potential distributions in N and F domains are determined from the electric current conservation and spin current re\-la\-xa\-ti\-on equations. In the dc regime they read \cite{rashba_epjb2002}
\begin{equation}
\label{eq:bulk_barmu_mus}
\Delta\mu = - \frac{p}{2} \Delta\mu^s \ , \quad
\lambda_s^2 \Delta\mu^s = \mu^s \ ,
\end{equation}
with $\lambda_s$ being the spin diffusion length, denoted as $\lambda_{sN}$ of $\lambda_{sF}$ in the corresponding materials.

The tunnel barrier between N and F is necessarily spin-selective, with unequal conductances $\Sigma_{\uparrow} \neq \Sigma_{\downarrow}$.\cite{rashba_prbrc2000} The current densities through the barrier are
\begin{eqnarray}
\label{eq:jtunnel}
j^{e}_{\perp} &=& - \frac{\Sigma}{e^2} \, ([\mu] + \frac{\Pi}{2}[\mu^s]) ,
\\
\label{eq:jstunnel}
j^s_{\perp} &=& -\frac{\Sigma}{2 e^2} \, ([\mu^s] + 2 \Pi [\mu]) ,
\end{eqnarray}
where $\Sigma = \Sigma_{\uparrow} + \Sigma_{\downarrow}$, $\Pi = (\Sigma_{\uparrow}-\Sigma_{\downarrow})/\Sigma$, and $[\mu] = \mu_F - \mu_N$, $[\mu^s]  = \mu^s_F - \mu^s_N$ are the potential jumps across the barrier. Eqs.~(\ref{eq:jtunnel}) and (\ref{eq:jstunnel}) provide the boundary conditions at the F/N interface.

In terms of spin potential $\mu^{s}$ right beneath the contact F$_2$, and the conductivity polarizations $p$ and $\Pi$, the Johnson-Silsbee formula states
\begin{equation}
\label{eq:JS_ohmic}
V = p \mu^s/2e \ ,
\end{equation}
for high-transparency Ohmic contacts, $\Sigma \to \infty$,\cite{johnson-silsbee_prl1985} and
\begin{equation}
\label{eq:JS_tunnel}
V = \Pi \mu^s/2e
\end{equation}
for a low-tranparency tunnel contacts, $\Sigma \to 0$.\cite{rashba_prbrc2000, rashba_epjb2002}

\subsection{Non-local voltage in the limit of large tunnel resistance}
\begin{figure}[t]
\center
\includegraphics[width = 0.45\textwidth]{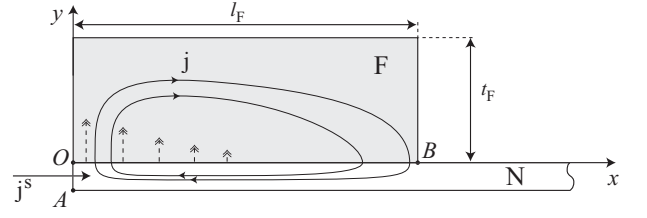}
    \caption{Right side of NLSV with pure spin current injection. Solid loops show the induced electric currents. Vertical dashed lines with double arrows represent the effective electromotive forces generated at the boundary due to spin accumulation.}
 \label{fig:domain}
\end{figure}

We consider now the right hand side of the NLSV (Fig.~\ref{fig:domain}). Point $O$ is the origin of the $(x,y)$ coordinate system. It is assumed that spin current is uniformly injected along the cross-section $OA$ of the normal line. The tunnel barrier se\-pa\-ra\-tes N and F along the segment $OB$ of length $l_F$. The normal wire N is considered to be infinitely long. Accordingly, Eqs.~(\ref{eq:bulk_barmu_mus}) have to be solved in the N and F domains with the following boundary conditions. Along the segment $OA$, $\mu^s(0)$ is a given constant, and the electric current component normal to $OA$ is zero. At all other outer boundaries, the normal components of both electric and spin currents vanish. At the segment $OB$ Eqs.~(\ref{eq:jtunnel}) and (\ref{eq:jstunnel}) relate the vertical current components to potential jumps.

For an infinitely high tunnel barrier one has $\Sigma_{\uparrow,\downarrow} = 0$, hence there is no penetration of spins into the F layer. The problem reduces to that of spin diffusion along the N line. With uniform spin injection along $OA$, the solution reads
\begin{equation}\label{eq:mus0}
\mu^s_0(x,y,N) = \mu^s(0) \exp(-x/\lambda_{sN})
\end{equation}
in N and $\mu^s_0(x,y,F) = 0$ in F. Furthermore, since the barrier is impenetrable, we essentially deal with two electrically disconnected conductors. Their electrochemical potentials are thus uniform $\mu_0(x,y) = \mu_{0}(N), \mu_{0}(F)$ and may assume arbitrary values.

For non-zero but small barrier conductivities given by $\Sigma_{\uparrow,\downarrow} = \varepsilon \tilde\Sigma_{\uparrow,\downarrow}$ with $\varepsilon \to 0$ we seek potentials in the form of Taylor expansion in $\varepsilon$
\begin{eqnarray*}
\mu(x,y,D) &=& \mu_0(D) + \varepsilon\mu_1(x,y,D) + \ldots \ ,
 \\
\mu^s(x,y,D) &=& \delta_{DN} \ \mu^s(0) e^{-x/\lambda_{sN}} + \varepsilon\mu^s_1(x,y,D) + \ldots \ ,
\end{eqnarray*}
where index $D = F,N$ defines the domain and $\delta_{DN}$ is the Kronecker symbol.

Finding the current densities in N and F requires calculating $\mu_1(x,y,D)$ and $\mu^s_1(x,y,D)$. However, to first order in $\varepsilon$ the non-local voltage can be found  without the full solution. From Eq.~(\ref{eq:jtunnel}), the leading term of the particle current across the barrier is first order in $\varepsilon$:
$$
j^e_{y}(x,0) = -\frac{\epsilon\tilde\Sigma}{e^2} \, \left( (\mu_{0F} - \mu_{0N}) + \frac{\Pi}{2} (-\mu^s_0(x,0,N)) \right) + \ldots
$$
Particle current conservation requires that the integral $\int_0^{l_F} j^e_y (x,0) dx$ vanish in the stationary state considered here. To first order in $\varepsilon$ this yields
\begin{equation}
\label{eq:averaging}
(\mu_{0F} - \mu_{0N}) l_F - \frac{\Pi}{2} \int_0^{l_F} \mu^s_0(x,0,N) dx = 0.
\end{equation}
Substituting the zeroth-order solution $\mu^s_0(x,0,N)$ from Eq.~(\ref{eq:mus0}), we find the measured voltage
$$
V = \frac{\mu_{0F} - \mu_{0N}}{e} = \frac{\Pi \lambda_{sN} (1 - e^{-l_F/\lambda_{sN}})}{2 e l_F} \mu^s(0).
$$
In the limit $l_F \ll \lambda_{sN}$, i.e., when spin accumulation under F is nearly constant, the tunnel Johnson-Silsbee result (\ref{eq:JS_tunnel}) is re\-co\-ve\-red. We can now present the voltage drop across the device of an arbitrary width $l_F$ as
\begin{equation}
\label{eq:voltage_wide_contact}
V(l_F) = \frac{\lambda_{sN}}{l_F} \left(1 - e^{-l_F/\lambda_{sN}} \right) V_{JS}.
\end{equation}
Eq.~(\ref{eq:voltage_wide_contact}) is the central result of our paper. In terms of the ``local Johnson-Silsbee voltage'' defined as $V_{JS}(x) = \Pi \mu_s(x,0,N)/(2 e)$, Eqs. (\ref{eq:averaging}) and (\ref{eq:voltage_wide_contact}) can be viewed as averaging $V_{JS}(x)$ over the contact width. However, as the contact width $l_F$ grows, electrochemical potentials in the F and N contacts become significantly non-uniform, and the simple picture above breaks down, as shown in the next section.

\subsection{Validity conditions for the large tunnel resistance approximation}
Equations (\ref{eq:j}) and (\ref{eq:jtunnel}) can be interpreted by taking the point of view that the particle current is produced not only by the electrochemical potential gradients, but also by an additional ``effective'' electromotive force (EMF) associated with non-uniformity of spin potential.\cite{rashba_prbrc2000, rashba_epjb2002, fabian_APS2007} Such a view helps one to visualize the emergence of circular currents.\cite{bazaliy:apl2017} Here we will use it to find the validity range of the approximation of the preceding section, that allowed us to neglect the variations of the electrochemical potentials in the N and F films.

\begin{figure}[t]
\center
\includegraphics[width = 0.45\textwidth]{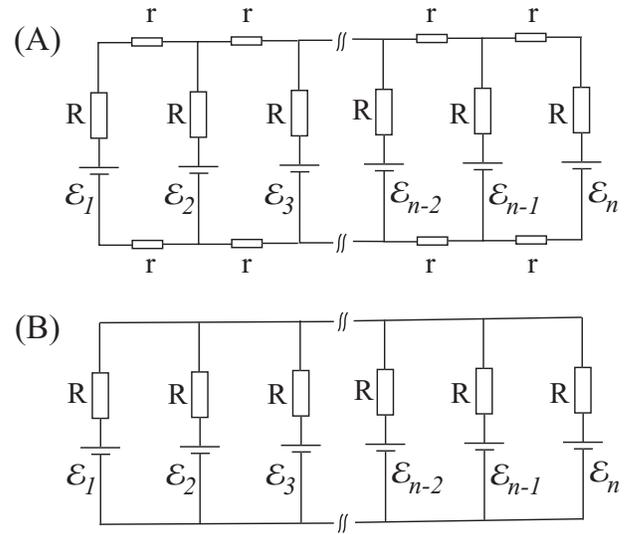}
    \caption{Electric circuits illustrating the emergence of circular current. Effective EMF's $\mathcal{E}_i$ represent the forces produced by spin potential imbalance. Circuit (A) takes into account the distributed resistance $r$ of the bulk N and F domains. Circuit (B) neglects $r$ as small compared to the resistance $R$ of the tunnel barrier.}
 \label{fig:circuit_analogy}
\end{figure}

For our device, the effective EMF interpretation leads to an electric circuit analogy, shown in Fig.~\ref{fig:circuit_analogy}A. Here $\mathcal{E}_i$ represent the effective EMFs, developing across the tunnel barrier due to the jump of $\mu^s$ as per Eq. (\ref{eq:jtunnel}). In the $\varepsilon \to 0$ limit this jump produces the leading, zeroth order contribution to effective EMF, while variation of $\mu_s$ in the ferromagnet brings first order corrections in $\varepsilon$. As one moves to the right, away from the spin injection cross-section $OA$, the jump $[ \mu_s ] = \mu_s(x,0,N) + \mathcal{O}(\varepsilon)$ decreases as per Eq.~(\ref{eq:mus0}), and the corresponding EMFs gradually decay to zero. Resistors $R$ represent the tunneling barrier resistance per unit length. Resistors $r$ represent the distributed resistance of the N and F layers.

Approximation (\ref{eq:voltage_wide_contact}) corresponds to neglecting the distributed resistance ($r$), which yields the circuit shown in Fig.~\ref{fig:circuit_analogy}B, where the upper and lower  horizontal lines are indeed characterized by constant electric potentials $V_F$ and $V_N$. Applying the Kirchhoff rules to Fig.~\ref{fig:circuit_analogy}B, one finds the voltage $V$ between F and N:
$$
V = V_F - V_N = \frac{\sum_{i = 1}^n \mathcal{E}_i}{n}
$$
where $i = 1, ..., n$ labels the EMFs, and $n$ is the total number of vertical legs. This equation is the analogue of the result (\ref{eq:voltage_wide_contact}).

Under what conditions can one ignore the resistors $r$ and replace Fig.~\ref{fig:circuit_analogy}A by Fig.~\ref{fig:circuit_analogy}B? Obviously, inequality $r \ll R$ has to be satisfied. This, however, is not enough: at the same time, the vol\-ta\-ge drop $\delta V$ along the upper and lower horizontal lines must be much smaller than $V$.

Horizontal voltage drop $\delta V$ can be expanded in powers of $r/R \ll 1$, and the leading term, linear in $r/R$, can be explicitly obtained in terms of the distribution of $\mathcal{E}_i$ (see Appendix). In our problem the $\mathcal{E}_i$ decay with increasing $i$, reflecting the decay of $\mu^s_0(x,N)$ along the $x$-axis. The Appendix shows that the ratio $\delta V/ V$ increases as the decay becomes more rapid, and only a few first EMFs remain non-zero---that is, as the contact length $l_F$ becomes large compared with the decay length $\lambda_{sN}$. This limit constrains $r/R$ the most stringently, as expressed by the inequality
$$r \ll \frac{R}{n^2} \ .$$
Such a condition shows, in particular, that as the number of vertical legs increases while $r/R$ is kept fixed, the approximation eventually breaks down.

For $R$ representing the tunnel barrier resistance per length $\Delta x$, continuous description is recovered by the correspondence
$$
R \to (\Sigma \Delta x)^{-1} \ ,
\quad
r \to (\sigma t/\Delta x)^{-1} \ ,
 \quad
n \to \frac{l_F}{\Delta x} \ .
$$
Here $\sigma t$ stands for either $\sigma_F t_F$ of $\sigma_N t_N$ since we assumed that both N and F lines can be described by the same $r$. If electric properties of the lines differ by orders of magnitude, a more involved analysis is required.

Together with the inequality $r \ll R/n^2$, this yields the condition
\begin{equation}\label{eq:validity_condition}
\Sigma \ll \frac{\sigma t}{l_F^2}.
\end{equation}

Inequality (\ref{eq:validity_condition}) can be rewritten as $\Sigma l_F \ll \sigma t/l_F$. Here the left hand side is the total vertical conductance of the tunnel barrier, and the right hand side is the total conductance of the F or N layer in the horizontal direction (more precisely, these are the conductances per unit depth of the device in the direction perpendicular to the plane of Fig.~\ref{fig:domain}).

\section{Conclusions}
Expression (\ref{eq:voltage_wide_contact}) shows that the non-local voltage measured by a tunneling F/N contact can strongly depend on the contact width. Note that in the limit of  low barrier conductance the suppression of voltage is independent of $\Sigma$. The latter can be very small, making the circular electric current behind the effect completely negligible. And yet, this current will significantly suppress the non-local voltage.

It is instructive to compare the evolution of non-local vol\-ta\-ge in transparent and tunnel barriers contacts. In the former case \cite{bazaliy:apl2017} there are two regimes: for $l_F \ll \lambda_{sN}$ the voltage is given by $V_{JS} = p \mu_s/2e$ (\ref{eq:JS_ohmic}), and is independent of voltmeter probe positions; for $l_f \geq \lambda_{sN}$ the voltage becomes probe-position dependent, and decreases compared with  (\ref{eq:JS_ohmic}). In the tunnel contact case there are three regimes: for $l_F \ll \lambda_{sN}$ the voltage is given by $V_{JS} = \Pi \mu_s/2e$ (\ref{eq:JS_tunnel}), independently of the probe positions; for $\lambda_s \leq l_F \ll \sqrt{\sigma t/\Sigma}$ the voltage is still independent of probe positions but reduces to the value (\ref{eq:voltage_wide_contact}); finally, for $l_F \geq \sqrt{\sigma t/\Sigma}$ the voltage becomes probe-position dependent, while being further reduced.

\section{Acknowledgements}
Ya.\ B.  is grateful to the Laboratoire de Physique Th\'eorique, Toulouse, for the hospitality, and to CNRS for funding the visits. R. R. thanks the Department of Physics \& Astronomy for the kind hospitality and support of his visit.

\appendix

\section{Calculation of the longitudinal voltage}

\begin{figure}[t]
\center
\includegraphics[width = 0.25\textwidth]{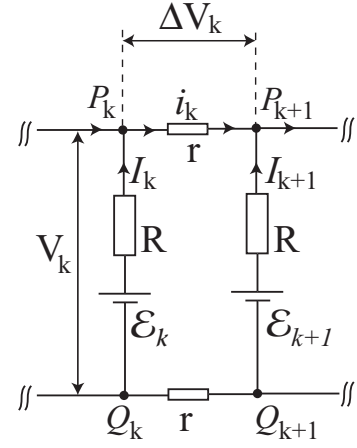}
    \caption{A unit of the effective circuit.}
 \label{fig:circuit_unit}
\end{figure}

An elementary unit of the original circuit is shown in Fig.~\ref{fig:circuit_unit}. Vertical legs are numbered by index $k = 1, 2, \ldots, n$. Current through a vertical leg $k$ is related to voltage $V_k$ between points $P_k$ and $Q_k$ as $I_k = (\mathcal{E}_k - V_k)/R$. Due to the symmetry between upper and lower lines, the voltage drop between points $P_k$ and $P_{k+1}$ is $\Delta V_k = (V_k - V_{k+1})/2$. Current through a horizontal leg connecting points $P_k$ and $P_{k+1}$ is then $i_k = \Delta V_k/r$. Current conservation at point $P_k$ gives
$$
i_k - i_{k-1} = \frac{\mathcal{E}_k - V_k}{R},
$$
or equivalently
\begin{equation}\label{eq:DeltaV_iteration}
\Delta V_k = \Delta V_{k-1} + \frac{r}{R} (\mathcal{E}_k - V_k).
\end{equation}
It's easy to check that at the left and right ends of the circuit we have to set  $\Delta V_0 = 0$ and $\Delta V_n = 0$. This will account for the fact that no current is entering the point $P_1$ from the left or leaving the point $P_n
$ to the right.

First, we express the condition that the sum of currents entering the upper line has to be zero
$$
\sum_{k = 1}^{n} I_k = \frac{1}{R}\sum_{k = 1}^{n} (\mathcal{E}_k - V_k) = 0 \ ,
$$
or
$$
\sum_{k = 1}^{n} V_k = \sum_{k = 1}^{n} \mathcal{E}_k \ .
$$
The average vertical voltage between the upper and lower lines, $\bar V = (\sum V_k)/n$ is then
\begin{equation}\label{eq:vertical_average}
\bar V = \frac{\sum_{k = 1}^{n} \mathcal{E}_k}{n}.
\end{equation}

Next, we wish to express the horizontal voltage drop $\delta V$ between the points $P_1$ and $P_n$, so that we can later require $\delta V \ll \bar V$ to make the average voltage a meaningful quantity. Using (\ref{eq:DeltaV_iteration}) and $\Delta V_0 = 0$ we can write
\begin{eqnarray*}
\Delta V_1 &=& \frac{r}{R} (\mathcal{E}_1 - V_1) \ ,
 \\
\Delta V_2 &=& \Delta V_1 + \frac{r}{R} (\mathcal{E}_2 - V_2) =
  \frac{r}{R} \left( (\mathcal{E}_1 - V_1) + (\mathcal{E}_2 - V_2)
  \right) \ ,
 \\
\ldots &&
 \\
\Delta V_k &=&
  \frac{r}{R} \left( (\mathcal{E}_1 - V_1) + (\mathcal{E}_2 - V_2) + \ldots
  + (\mathcal{E}_k - V_k)
  \right) \ .
\end{eqnarray*}
The total horizontal voltage drop between $P_1$ and $P_n$ can be then expressed by summing these voltage drops
\begin{eqnarray*}
\delta V &=& \sum_{k = 1}^{n-1} \Delta V_k =
\frac{r}{R} \big[ (n-1)(\mathcal{E}_1 - V_1) + (n-2)(\mathcal{E}_2 - V_2) + \ldots
\\
&&
  + (n-k)(\mathcal{E}_k - V_k) + \dots + (\mathcal{E}_{n-1} - V_{n-1})
  \big].
\end{eqnarray*}
Clearly, for $r = 0$ one gets $\delta V = 0$ and $V_k = \bar V$ for all $k$'s. For non-zero $r$ we can consider an expansion in powers of $r/R \ll 1$
$$
V_k = \bar V + \left( \frac{r}{R} \right) V_{k1} + \left( \frac{r}{R} \right)^2 V_{k2} + \ldots
$$
Then in the first order in $r/R$ the total horizontal voltage drop is
\begin{equation}\label{eq:deltaV_first_form}
\delta V \approx \frac{r}{R} \sum_{k = 1}^{n-1} (n-k)(\mathcal{E}_k - \bar V).
\end{equation}
This is an explicit formula for $\delta V$ in terms of the given set of $\mathcal{E}_k$'s. Note that it excludes the leftmost EMF $\mathcal{E}_n$, however, using the identity
$\sum_{k=1}^n (\mathcal{E}_k - \bar V) =0$ that follows from (\ref{eq:vertical_average}), and adding it to (\ref{eq:deltaV_first_form}), one can rewrite it so that all EMF's are present on equal footing
\begin{equation}\label{eq:deltaV}
\delta V \approx \frac{r}{R} \sum_{k = 1}^{n} (n +1 -k)(\mathcal{E}_k - \bar V).
\end{equation}
Total horizontal voltage drop $\delta V$ depends on the distribution of EMF's. For example, in the absence of spatial variation, $\mathcal{E}_k = {\rm const}$, Eq.~(\ref{eq:deltaV}) gives $\delta V = 0$, regardless of the value of $r$.

The more spatial variation of EMF's there is, the larger becomes $\delta V$. This can be illustrated by an example, where the first $p<n$ EMFs are equal and non-zero, $\mathcal{E}_1 = \mathcal{E}_2 = \ldots = \mathcal{E}_{p} = \mathcal{E}$, while all the following EMFs vanish: $\mathcal{E}_{p+1} = \mathcal{E}_{p+2} = \ldots = \mathcal{E}_{n} = 0$. For such an EMF distribution the average voltage is $\bar V = (p/n) \mathcal{E}$, and Eq. (\ref{eq:deltaV}) leads---after some algebra---to
$$
\delta V = \frac{r}{R} \frac{n(n-p)}{2} \bar V \ .
$$
Whenever $p$ is small enough compared with $n$, so that $(n-p) \sim n$, this formula yields
$
\delta V  \sim \bar V n^2 r / R  \ ,
$
and condition $\delta V \ll \bar V$ then leads to a requirement
\begin{equation}\label{eq:appendix:conclusion}
r \ll \frac{R}{n^2} \ .
\end{equation}

\end{document}

\begin{center}
{\Huge Supplement}\\
\bigskip
to Ya.\ B. Bazaliy and R. R. Ramazashvili, \\
``Closed-loop electric currents and non-local resistance measurements with wide F/I/N tunnel contacts''.
\end{center}

\section*{Notation for spin and charge diffusion equations}

In the Valet-Fert model carrier distributions for spin $\alpha = \, \uparrow,\downarrow$ are characterized by different electrochemical potentials $\mu_{\alpha}$. Currents ${\bf j}_{\sigma}$ (vectors in real space) are defined here as particle number currents. To obtain electric currents, they should be multiplyed by electron charge. With two conductivities $\sigma_{\uparrow, \downarrow}$ being different in a ferromagnet, the currents carried by the two spin populations are given by ${\bf j}_{\alpha} = - (\sigma_{\alpha}/e^2) \nabla \mu_{\alpha}$.

Electric current ${\bf j} = {\bf j}_{\uparrow} + {\bf j}_{\downarrow}$ is conserved and the total electron density $n = n_{\uparrow} + n_{\downarrow}$ obeys
$$
\partial_t n + {\rm div} {\bf j} = 0 \ .
$$
The spin current ${\bf j}^s = {\bf j}_{\uparrow} - {\bf j}_{\downarrow}$ is not conserved due to the spontaneous relaxation of spin.

\bigskip

In the normal metal $\sigma_{\uparrow} = \sigma_{\downarrow}$, and  in equilibrium $n_{\uparrow} = n_{\downarrow}$. For small nonequilibrium spin density $n_s = n_{\uparrow} - n_{\downarrow}$,  spin decay is characterized by a relaxation time $\tau_s$, so that spin current and spin density are related by
$$
\partial_t n^s + {\rm div} {\bf j}^s = - \frac{n_s}{\tau_s} \ .
$$
In a steady state one finds
$$
{\rm div}{\bf j} = 0, \quad
    {\rm div}{\bf j}^s = - n_s/\tau_s \ .
$$
Particle and spin currents in a normal metal can be expressed through the average potential $\mu = (\mu_{\uparrow} + \mu_{\downarrow})/2$ (the quantity measured by an ideal voltmeter), and the spin potential
$\mu^s~=~\mu_{\uparrow}~-~\mu_{\downarrow}$ that characterizes the non-equilibrium spin accumulation. With $\sigma = \sigma_{\uparrow} + \sigma_{\downarrow}$ one finds
$$
{\bf j} = - \frac{\sigma}{e^2} \, \nabla\mu \ ,
\qquad
{\bf j}^s = -\frac{\sigma}{2 e^2} \, \nabla\mu^s \ .
$$
Spin accumulation and spin potential are related by the density of states $\nu$ as $n_s = \nu \mu^s$, giving
\begin{equation}\label{eq:divj_divjs}
{\rm div}{\bf j} = 0, \quad
    {\rm div}{\bf j}^s = - \nu \mu^s/\tau_s \ .
\end{equation}
or, assuming spatially uniform material parameters,
\begin{equation}\label{eq:nablamu}
\Delta \mu = 0 \ , \qquad \Delta \mu^s = \frac{\mu^s}{\lambda_s^2}
\end{equation}
with spin diffusion length $\lambda_s = \sqrt{\sigma \tau_s/(2 \nu e^2)}$.

\bigskip

In a ferromagnet $\sigma_{\uparrow} \neq \sigma_{\downarrow}$, and the currents ${\bf j}$ and ${\bf j}^s$ can be written as
\begin{eqnarray}
\label{eq:j}
{\bf j} &=& - \frac{\sigma}{e^2} \, (\nabla\mu + \frac{p}{2}\nabla\mu^s)
\\
\label{eq:js}
{\bf j}^s &=& -\frac{\sigma}{2 e^2} \, (\nabla\mu^s + 2 p \nabla\mu)
\end{eqnarray}
with polarization $p~=~(\sigma_{\uparrow}~-~\sigma_{\downarrow})/\sigma$. Note that in the Eqs. (\ref{eq:j}-\ref{eq:js}) spin and charge are coupled by $p \neq 0$. Spin relaxation description in a ferromagnet is more complicated and involves two spin-dependent densities of states $\nu_{\uparrow}$, $\nu_{\downarrow}$.\footnote{E. I. Rashba, Eur. Phys. J. B {\bf 29}, 513 (2002) {\em Diffusion theory of spin injection through resistive contacts}.} The steady state equations acquire the form
\begin{equation}
\label{eq:bulk_barmu_mus}
\Delta\mu = - \frac{p}{2} \Delta\mu^s \ , \quad
\lambda_s^2 \Delta\mu^s = \mu^s \ ,
\end{equation}
with $\lambda_s$ being the appropriately defined$^1$ spin relaxation length in a ferromagnet. Note that the second equation on $\mu_s$ does not involve the electric potential $\mu$ for either normal metal of ferromagnet. However, the two are coupled for $p \neq 0$ by the first equation.